\documentclass[12pt]{article}

\usepackage{scicite}

\usepackage{times}
\usepackage{setspace}
\usepackage{graphicx}

\newenvironment{sciabstract}{%
\begin{quote} \bf}
{\end{quote}}

\newcounter{lastnote}
\newenvironment{scilastnote}{%
\setcounter{lastnote}{\value{enumiv}}%
\addtocounter{lastnote}{+1}%
\begin{list}%
{\arabic{lastnote}.}
{\setlength{\leftmargin}{.22in}}
{\setlength{\labelsep}{.5em}}}
{\end{list}}

\title{A High Phase-Space-Density Gas of Polar Molecules}

\author{K.-K. Ni,$^{1\ast}$ S. Ospelkaus,$^{1\ast}$ M. H. G. de Miranda,$^{1}$ A. Pe'er,$^{1}$\\B. Neyenhuis,$^{1}$
 J. J. Zirbel,$^{1}$ S. Kotochigova,$^{2}$ P. S. Julienne,$^{3}$\\
 D. S. Jin,$^{1\dagger}$ J. Ye$^{1\dagger}$\\
\\
\normalsize{$^{1}$JILA, NIST and University of Colorado,}\\
\normalsize{Department of Physics, University of Colorado, Boulder}\\
\normalsize{Boulder, CO 80309-0440, USA}\\
\normalsize{$^{2}$Physics Department, Temple University, Philadelphia, PA 19122-6082, USA}\\
\normalsize{$^{3}$Joint Quantum Institute, NIST and University of Maryland, Gaithersburg,}\\
\normalsize{Gaithersburg, MD 20899-8423, USA}\\
\\
\normalsize{$^\ast$These authors contributed equally to this work; }\\
\normalsize{$^\dagger$To whom correspondence should be addressed; }\\
\normalsize{E-mail:  jin@jilau1.colorado.edu; ye@jila.colorado.edu}
}

\date{}

\begin{document} 


\baselineskip24pt


\maketitle


\begin{sciabstract}
A quantum gas of ultracold polar molecules, with long-range and anisotropic 
interactions, would not only enable explorations of a large class
of many-body physics phenomena, but could also be used for
quantum information processing. We report on the creation of an
ultracold dense gas of $\,^{40}$K$^{87}$Rb polar molecules. Using a
single step of STIRAP (STImulated Raman Adiabatic Passage) via two-frequency 
laser irradiation, we
coherently transfer extremely weakly bound KRb molecules to the
rovibrational ground state of either the triplet or the singlet electronic ground
molecular potential. The polar molecular gas has a peak density of 10$^{12}$
cm$^{-3}$, and an expansion-determined translational temperature of 350 nK.
The polar molecules
have a permanent electric dipole moment, which we measure via Stark
spectroscopy to be 0.052(2) Debye for the triplet rovibrational ground
state and 0.566(17) Debye for the singlet rovibrational ground state.
(1 Debye= 3.336$ \times 10^{-30}$ C m)
\end{sciabstract}


Ultracold atomic gases have enjoyed tremendous success as model
quantum systems in which one can precisely control the particles'
internal degrees of freedom and external motional states. These
gases make interesting many-body quantum systems when the effects of
interactions between the particles, along with their quantum
statistics, are important in determining the macroscopic response of the system.
However, for most atomic gases the interactions are exceedingly
simple:  they are spatially isotropic and are sufficiently
short-range to be well approximated by contact interactions.  A
wider range of many-body physics phenomena could be explored if the gas comprised
particles with more complex interactions, such as would occur in
an ultracold gas of polar molecules.  Here, the electric
dipole-dipole interaction is long-range and spatially anisotropic,
much like the interaction of magnetic spins in condensed matter
systems.  Dipole-dipole interactions can be realized
using atomic magnetic dipoles~\cite{cr2005, cr2007}, but are
typically much weaker than those that could be
realized for molecules with a permanent electric dipole moment.
Theoretical proposals employing ultracold polar molecules range from
the study of quantum phase transitions~\cite{zoller} and 
quantum gas dynamics~\cite{dynamic} to
quantum simulations of condensed matter spin systems\cite{zollerCM} and schemes 
for quantum information processing ~\cite{demille, andre, yelin}.

The relative strength of dipole-dipole interactions in an ultracold
gas depends critically on three parameters - the temperature T, the
dipole moment, and the number density of molecules in the sample.  For interaction
effects to be strongly manifested, the interaction energy must be comparable to or
greater than the thermal energy.  This condition calls for low temperatures
and large dipole moments.  In addition, a high number
density is needed since the dipole-dipole interaction scales as
$1/R^{3}$, where $R$ is the distance between particles.  The combined
requirements of low temperature and high density can only be met if 
the molecule gas has a high phase-space density, i.e.
the gas should be near quantum degeneracy.  Recently, there has been
rapid progress toward creating samples of cold polar molecules
~\cite{sageRbCs, wangKRb, buffer, stark, stark2}; however, it remains a
challenge to create a gas where dipole-dipole interactions are observable.  

Direct cooling of ground-state molecules
~\cite{buffer, stark, stark2} has thus far only attained milliKelvin final temperatures. 
An alternative route is to start with a high
phase-space-density gas of atoms and then coherently and efficiently convert
atom pairs into ground-state molecules without heating the sample
~\cite{heinzen2000, RbSTIRAP, sospelkaus2008, cs2}.
To create polar molecules, the initial atomic gas must be
a mixture of two types of atoms so that the resulting
diatomic molecules are heteronuclear. In
addition, only tightly bound molecules will have an appreciable
electric dipole moment.  This requirement gives rise to the considerable
challenge of efficiently converting atoms that are relatively far
apart into molecules of small internuclear distance, without allowing the released binding
energy to heat the gas.  

Preserving the high phase space density of the initial gas while
transferring atoms to deeply bound polar
molecules requires coherent state transfer.  
Here we report the efficient transfer of ultracold atoms into the 
rovibrational ground state of both the triplet and the singlet
electronic ground molecular potentials,
and a measurement of the resulting molecules' electric dipole moments.
We accomplish this goal by creating near-threshold molecules and then
using a single step of STImulated Raman Adiabatic Passage (STIRAP)~\cite{STIRAP}.
Key steps in realizing efficient transfer with STIRAP are the identification of a favorable
intermediate state and the ability to maintain phase coherence of
the Raman lasers. With the coherent transfer to a single quantum state, we 
create $3\cdot10^4$ rovibrational ground-state polar molecules at 
a peak density of 10$^{12}$ cm$^{-3}$.
The molecules are created in an optical dipole trap and their expansion energy
is $k_b\cdot350$ nK, where $k_b$ is the Boltzmann's constant.


The starting point for this work is a near quantum degenerate gas
mixture of fermionic $^{40}$K atoms and bosonic $^{87}$Rb atoms
confined in an optical dipole trap.  We use a magnetic-field tunable
Fano-Feshbach resonance at 546.7 G~\cite{KRbFeshbach, lens} 
to associate atoms into extremely weakly bound diatomic molecules\cite{fmreview, KRblattice}.
With an adiabatic
magnetic-field sweep across the resonance, we typically create
a few $10^4$ near-threshold molecules. The experiments described here
are performed at a magnetic field of 545.9 G, where the
Feshbach molecules have a binding energy of $h\times$230 kHz 
($h$ is the Planck's constant).  We directly detect the Feshbach molecules
using time-of-flight absorption imaging. Given the measured number,
trap frequency, and expansion energy of the fermionic molecules, we
find that the Feshbach molecule gas is nearly quantum degenerate with
$T/T_F\approx$2, where $T_F$ is the Fermi temperature. Details of
the Feshbach molecule creation and detection have been described
elsewhere ~\cite{zirbel2008, zirbel2008pra}.

For efficient transfer from the initial state of Feshbach molecules to the final
state of tightly bound ground-state molecules, the chosen intermediate,
electronically excited state must have favorable wave
function overlap with both the initial and final states. 
We first demonstrate coherent transfer of the Feshbach molecules to the
rovibrational ground state of the triplet electronic ground potential, $a^3\Sigma$.
This state lies about 4000 cm$^{-1}$ above the absolute ground state, which
is the rovibrational ground state of the singlet electronic ground potential, $X^1\Sigma$.
Triplet and singlet refer to a total electronic spin of the molecule
that is one or zero, and our Feshbach molecules are predominately triplet in character.

\section*{Scheme for Transfer of Feshbach Molecules to $a^3\Sigma$ $v=0$ Molecules}

Our transfer scheme (Fig. 1) involves three molecular levels, the initial state $|i\rangle$, 
the intermediate 
state $|e\rangle$, and the final state $|g\rangle$, that are coupled by two laser fields, $\Omega_1$
 and $\Omega_2$. The first laser field, $\Omega_1$, drives
the up transition where the wave function amplitude of the weakly
bound Feshbach molecule state ($|i\rangle$) overlaps favorably near the Condon point
 to a deeply-bound
vibrational level, the tenth vibrational level, $v' = 10$, of the electronically excited $2^3\Sigma$
potential ($|e\rangle$).  The Condon point is the internuclear distance 
where the photon energy matches the difference between the excited and ground-state
potential energy curves.
The second laser field, $\Omega_2$, drives the down transition, with the Condon point  
near the outer turning point of the $v' = 10$ state, where its wave function overlaps strongly with
 the wave function for the ground vibrational level of the electronic ground $a^3\Sigma$ 
 potential ($|g\rangle$).

The Raman system for the
coherent state transfer employs a diode laser and a Ti:Sapphire laser.
Both lasers have tunable wavelengths around 1 $\mu$m. The two
lasers are individually phase-locked to a femtosecond optical
frequency comb~\cite{comb}, which itself is referenced to a stable 1064 nm
Nd:YAG laser. The phase-locked linewidth is measured to be narrower
than 25 Hz~\cite{comb1hz, combnote}. The comb covers the spectral
range from 532 nm to 1100 nm,
with a mode spacing of 756 MHz. The large 
wavelength span and precision referencing capability of the comb
are well-suited for spectroscopy in search of previously unobserved states.


To find the intermediate state, we performed a search guided by \textit{ab initio} calculation fitted to 
experimental data from molecules of different isotopes,
$^{39}$K$^{85}$Rb~\cite{k39rb85}, with the appropriate mass scalings. We
located excited molecular states by measuring the loss of Feshbach
molecules as a function of the applied laser wavelength ($\Omega_1$). This laser 
excites a one-photon bound-bound transition that is followed by spontaneous decay into other states. 
We have observed the
$v'=8$ to $v'=12$ vibrational levels of the $2^3\Sigma$ excited
potential with a roughly 1300 GHz spacing between neighboring vibrational levels, and we chose
$v'=10$ as our intermediate state. From the measured one-photon loss
rate and power-broadened one-photon lineshapes, the transition dipole
 moment\cite{transitionDnote} is determined to be 0.004(2) $ea_0$ (1 $ea_0=2.54$ Debye$=8.48
 \cdot 10^{-30}$ C$\cdot$m).

\section*{The $a^3\Sigma$ $v=0$ Level}

To search for the triplet vibrational ground state $(a^3\Sigma,\,v=0)$,
we performed two-photon dark resonance spectroscopy\cite{STIRAP} in the limit of a
strong pump ($\Omega_2$) and weak probe ($\Omega_1$). Based on the
KRb potential published by Pashov \textit{et al.}~\cite{krbpotential}, we
calculated the triplet $v=0$ binding energy with a predicted
uncertainty of 0.1\%.  For the search, it was convenient to fix the frequency of
the weak probe laser to resonantly drive the transition from the
initial Feshbach molecule state to the $v'=10$ intermediate state.
The probe laser by itself causes complete loss of all the Feshbach
molecules.
We then varied the frequency of the strong coupling laser, and monitored
the initial state population after pulsing on both laser fields simultaneously.
When the Raman condition is fulfilled, i.e. 
the initial and final state energy splitting is matched by the
two laser frequency difference, the initial state population reappears (Fig. 2A).

The measured binding energy of the triplet $v=0$ molecules is $h\times$7.18
THz (corresponding to 240 cm$^{-1}$) at 545.94 G. We find that the $v=0$ level has rich hyperfine
plus rotational structure at this magnetic field (see Fig. 2A).
Because the accessible final states are influenced by
 selection rules, we have performed the two-photon
spectroscopy using two different states of the $v'=10$ intermediate
level. In addition to the triplet 
$v=0$ level, we have
also observed similar ground-state hyperfine structure for the
$v=1$ and $v=2$ levels of the $a^3\Sigma$ state, which have a vibrational energy spacing
of roughly 500 GHz, consistent with our theoretical prediction.

We have identified the quantum numbers of the three lowest
energy triplet $v=0$ states seen in the two-photon spectrum. The peaks labeled 1, 2, and 3 in Fig. 2A 
occur at binding energies of
$h\times$7.1804180(5) THz, $h\times$7.1776875(5) THz, and $h\times$7.1772630(5) THz, respectively. Peak
1 corresponds to the lowest hyperfine state in the rotational ground-state ($N=0$), peak 2
is a different hyperfine state with $N=0$, and peak 3 is the
lowest energy hyperfine state with $N=2$, where $N$ is the rotational quantum number.
This identification is based on Hund's coupling case (b),
where spin and molecular rotation are essentially decoupled and
the molecular hyperfine structure can be understood from
calculations using a separated atom basis with the rotational
progression appearing as a constant shift to all hyperfine levels.
Because of parity selection rules for optical transitions, we 
observe only states with an even $N$. The
calculated rotational constant is $B=0.5264$ GHz, which gives a predicted 
splitting between the $N=0$ and $N=2$ levels of $6B=3.158$ GHz.  

Using a dark resonance measurement such as shown in Fig. 2B, 
we have measured the strength of the $|e\rangle$ to $|g\rangle$ transition. 
Here, we
fix the down leg ($\Omega_2$) laser frequency and scan the up leg ($\Omega_1$) laser 
frequency. 
From the
width of the dark resonance for the rovibrational triplet
ground-state (peak 2), we find that we can drive the transition from
$v'=10$ to the triplet $v=0$ state with a Rabi frequency of 2$\pi \cdot8$ MHz. This
measurement used 60 $\mu$W of laser power focused to a beam waist
of $55$ $\mu$m.  The transition dipole moment derived from this
measurement is 0.20(2) $ea_0$, which is only one order of magnitude
weaker than a typical atomic optical transition.

\section*{High Phase-Space-Density Gas of Triplet Ground-State Polar Molecules}

We used peak 2 as the
rovibrational ground-state target for our coherent state transfer,
which is performed using the counter-intuitive pulse sequence of
STIRAP\cite{STIRAP}. The STIRAP beams are
co-propagating in order to minimize photon recoil.
The measured time evolution of the initial-state population
during a double STIRAP pulse sequence is shown in Fig. 3A. 
The roundtrip transfer efficiency of 31\% implies a 
one-way transfer efficiency of 56\%, which corresponds to
$3\cdot 10^4$ triplet $v=0$ $N=0$ polar molecules at a peak density of
$10^{12}$ cm$^{-3}$. Our transfer technique allows us to
reach a single quantum state without heating.
The expansion energy of the $v=0$ molecules
is measured after transferring them back to the Feshbach molecule state.
Using this expansion energy and the trap frequency
measured for the Feshbach molecules, the phase space density of the
polar molecule gas corresponds to $T/T_F\approx2.5$.

We measure the lifetime of the rovibrational ground-state molecules
by varying the hold time between transferring the molecules into the
$v=0$ $N=0$ state and bringing them back to the Feshbach molecule state for
imaging.  The lifetime is measured to be 170 $\mu$s as shown in Fig.
3B.  This lifetime may be limited by collisions 
with background atoms, which can induce
spin flips and cause molecules to decay into the lower lying singlet electronic ground potential.
The collisional decay could be reduced either by perfecting
the removal of the remaining atoms or by starting the molecule
production with atom pairs tightly confined in individual sites of an
optical lattice. The short lifetime of the final state currently limits the STIRAP transfer efficiency.

We demonstrate that KRb molecules in the triplet
rovibrational ground state are polar by directly measuring
their electric dipole moment. The predicted dipole moment for KRb
triplet rovibrational ground-state molecules is 0.05(3) Debye (D) ~\cite{kotochigova2003}. 
This is nine orders of magnitude
larger than the calculated $5\cdot 10^{-11}$ D dipole moment of
the initial Feshbach molecules and only about one order of magnitude smaller than a typical
polar molecule dipole moment of 1 Debye.  To measure the dipole moment, we
performed DC Stark spectroscopy on the three lowest energy states
observed in the two-photon spectrum (Fig. 2A).  We applied a DC electric field in the range from 0 to
2 kV/cm using a pair of indium tin oxide coated transparent electric field plates that are
separated by 1.3 cm outside the glass-cell based vacuum chamber. We measured the Stark
shift using the dark resonance spectroscopy discussed above. This
two-photon spectroscopy measures the energy splitting between the
initial and final states, and because the initial state has a
negligible dipole moment, any frequency shift of the dark resonance
can be attributed to the final-state Stark shift. For these
measurements we lowered the laser powers to give a dark resonance
width of 500 kHz.  The measured ground-state energies versus electric
field are shown in Fig. 4.

The effect of a DC electric field is to couple states of opposite
parity. For the $a^3\Sigma$ $v=0$ molecules, 
the opposite parity states are even-$N$ and odd-$N$
rotational states.  The two lowest energy states, which are 
the rotational ground state
$N=0$, exhibit similar Stark shifts.  From the measured Stark shift,
we find that the molecules' electric dipole moment is 0.052(2) D. 
The Stark shift of the third energy state, corresponding to
peak 3, is measured to be about 10 times smaller than that for the
peak 1 and peak 2 states.  This smaller Stark shift for the $N=2$ state is consistent with an
electric dipole moment of 0.052 D.

\section*{Absolute Ground-State Polar Molecules}

We have also achieved transfer of Feshbach molecules to the rovibrational
ground state of the singlet electronic ground potential, $X^1\Sigma$, thereby producing absolute
ground-state polar molecules. Reaching the singlet
rovibrational ground state requires the identification of a new suitable intermediate
state and the preservation of the phase-coherence of the Raman laser system over a much 
larger spectral difference.

A favorable scheme for transferring KRb Feshbach molecules, which have a
predominantly triplet character, 
to the rovibrational ground state of the singlet electronic potential, $X^1\Sigma$, requires 
an intermediate state that has strong electronic spin-orbit coupling\cite{stwalley2004, sageRbCs}, 
in addition to good wavefunction overlap with both the initial state and the target state.
The intermediate state we chose (Fig. 5A) is again associated with the excited potential $2^3\Sigma$.
This excited potential is split into two components labeled by $\Omega=1$ and $\Omega=0$,
corresponding to the total electronic angular momentum projection onto the molecular 
internuclear axis ($\Omega$).
The higher lying vibrational levels of the $\Omega=1$ component have strong triplet-singlet spin
mixing with the nearby excited singlet potential, $1^1\Pi$ \cite{bergeman}.
This spin
mixing provides the necessary coupling to transfer predominantly triplet character Feshbach molecules
to the singlet rovibrational ground state. In addition, these levels guarantee a large Franck-Condon
factor as the down leg Condon point is near the inner turning point 
of the intermediate state and coincides with the bottom 
of the $X^1\Sigma$ potential, while the up leg Condon point is
near the outer turning point of the intermediate state and also overlaps favorably
with the Feshbach molecule state. 

For our transfer scheme (Fig. 5A), we chose $v'=23$, $\Omega=1$ of the electronically
excited $2^3\Sigma$ potential as 
the intermediate state. We identified this state using the single-photon spectroscopy
technique discussed in the previous section.
We then proceeded with two-photon dark resonance spectroscopy
to search for the rovibrational ground state of the singlet electronic ground potential.
The two Raman lasers are
near 970 nm and 690 nm.
We measured the binding energy of the singlet rovibrational ground state ($v=0, N=0$ of
$X^1\Sigma$) to be $h\times$125.319703(1) THz (corresponding to about 0.5 eV) 
at 545.88 G. This deviates by only about 400 MHz (corresponding to 0.013 cm$^{-1}$) from our 
theoretical prediction based on the potential by Pashov \textit{et al.}\cite{krbpotential}.
We also located the rotationally excited $N=2$ state of the singlet vibrational
ground level at a binding energy of 125.313019(1) THz. The energy difference of the two 
states, 6$B=6.6836(5)$ GHz, gives a measured rotational constant $B=1.1139(1)$ GHz, which
agrees with the predicted value of 1.1140 GHz. 
From the dark resonance lineshape, we extract the
transition strengths for both the upward and the downward transitions and obtain 
 0.005(2) $ea_0$ and 0.012(3) $ea_0$, respectively.

Absolute ground-state KRb polar molecules are expected to have 
a much larger electric dipole moment than the triplet ground-state molecules. Many theoretical
efforts have calculated the dipole moment, with the results ranging from 0.5 D
to 1.2 D \cite{singletdipole}.
To measure the electric dipole moment, we have 
performed DC Stark spectroscopy on the $N=0$ and $N=2$ states of the singlet
vibrational ground level (Fig. 6). From the measured $N=0$ Stark shift, and the measured rotational 
constant $B$, we find that the singlet vibrational ground-state molecules have a permanent 
electric dipole moment of 0.566(17) Debye, which is $\sim$10
times larger than that of the triplet rovibrational ground state.
The measured Stark shift of the $N=2$ state is consistent with this value for the electric
dipole moment.
This large dipole moment allows the molecules to
be polarized by modest electric fields and will facilitate exploration of interaction effects.

Finally, we demonstrate the creation of absolute ground-state 
molecules ($v=0$, $N=0$ of $X^1\Sigma$)
using a single step of STIRAP. 
Using a STIRAP pulse length of 4$\mu$s for each transfer 
(Fig. 5B and 5C), we recovered
69\% of Feshbach molecules after a roundtrip transfer. The one-way transfer efficiency,
assuming equal efficiency on the two transfers, is 83\%. 
We also performed a roundtrip STIRAP pulse sequence with a 30 ms hold in between each
transfer, i.e. molecules are in the absolute ground state for 30 ms, and recovered 30\%
of the Feshbach molecules after the roundtrip transfer.
This shows that the absolute ground-state molecules are trapped
and have a much longer lifetime
than the triplet rovibrational ground-state molecules. 
We measured the expansion energy of the Feshbach molecules that were transferred back 
after a 20$\mu$s hold in the absolute ground state and observed no heating when comparing 
to the initial gas of Feshbach molecules. 
In the future, we anticipate
that near unity transfer efficiency should be possible with improved stabilization of the phase
coherence between the two Raman lasers. This ability to create a 
long-lived quantum gas
of ground-state polar molecules can be extended to other bialkali molecules and paves the 
way for future studies of dipolar Fermi gases and dipolar Bose-Einstein 
condensates\cite{ticknor2008, bohn,baranov}.



\bibliography{scibib}

\begin{thebibliography}{10}
\bibitem{combnote}
When the YAG is stabilized to an ultrastable laser, such as a strontium clock, the two-photon linewidth is narrower than 1 Hz.
\end{thebibliography}

\bibliographystyle{Science}


\begin{scilastnote}
\item This work has been supported by the NSF, NIST, AFOSR, and the W. M. Keck
Foundation. We thank D. Wang for 
experimental assistance and J. Bohn for discussion. K.-K. N.
and B. N. acknowledge support from the NSF,  S. O. from the
A.-v.\,Humboldt Foundation, M. H. G. de M. from the CAPES/Fulbright, and P. S. J. from
ONR.
\end{scilastnote}


\clearpage

\newpage

\begin{figure}[t]
\begin{center}
  \includegraphics[totalheight=13.0cm,width=11cm]{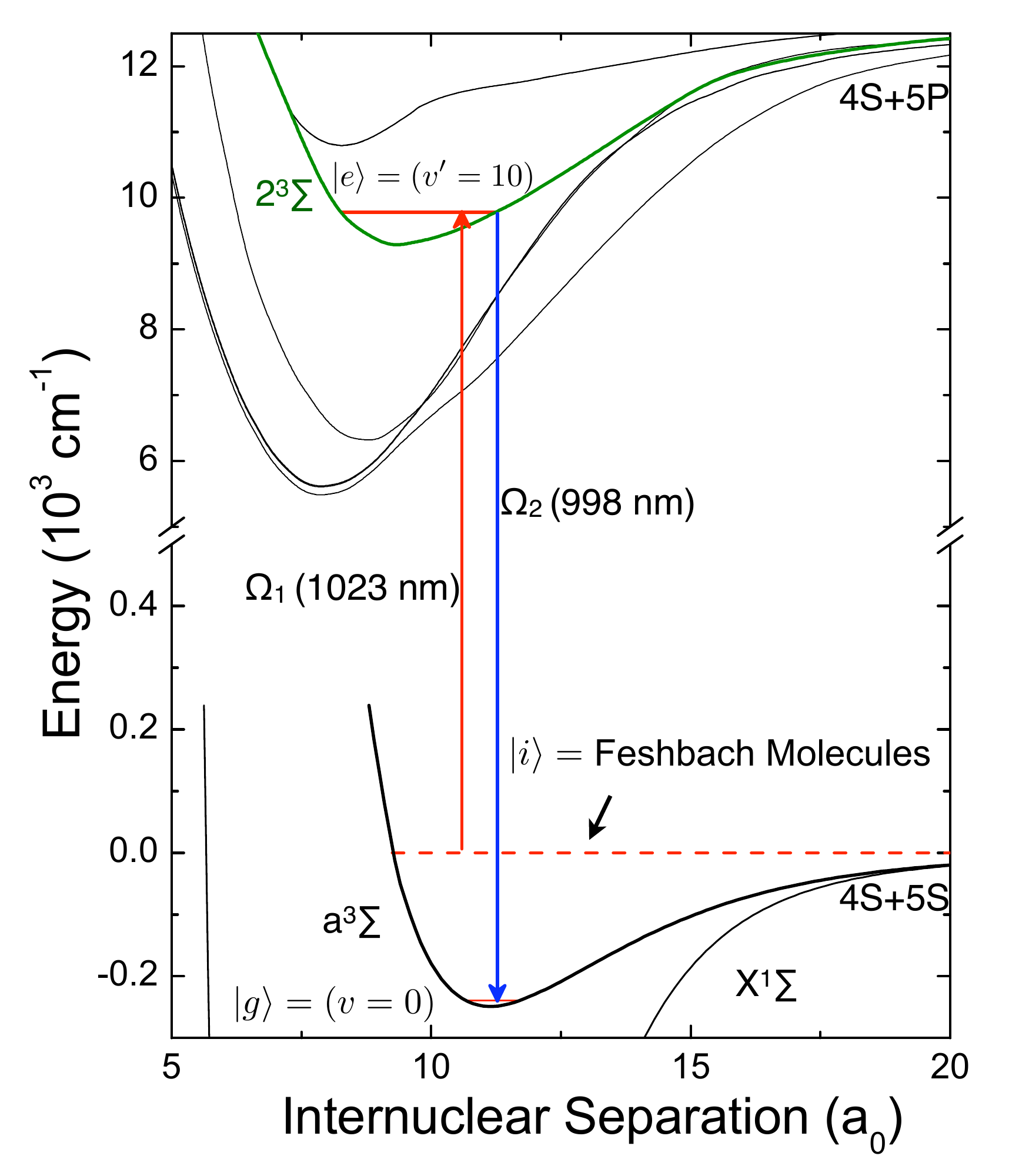}
   \caption{ \doublespacing Diagram of the KRb electronic ground and excited molecular potentials 
   and the vibrational levels
involved in the two-photon coherent state transfer to the triplet ground state.  Here, the
intermediate state $|e\rangle$ is the $v'=10$ level of the electronically
excited $2^3\Sigma$ potential.  The vertical arrows are placed at the respective Condon 
points of the up and down transitions. The intermediate state has favorable
transition dipole moments for both the up leg ($|i\rangle$ to $|e\rangle$) and
the down leg ($|e\rangle$ to $|g\rangle$), where the initial state $|i\rangle$ is a
weakly bound Feshbach molecule and the final state $|g\rangle$ is the
rovibrational ground state ($v=0$ $N=0$) of the triplet electronic ground potential, $a^3\Sigma$.
      \label{fig:1}}
\end{center}
\vspace{-0.05\textwidth}
\end{figure}

\newpage
\begin{figure}[t]
\begin{center}
   \includegraphics[totalheight=11cm,width=14cm]{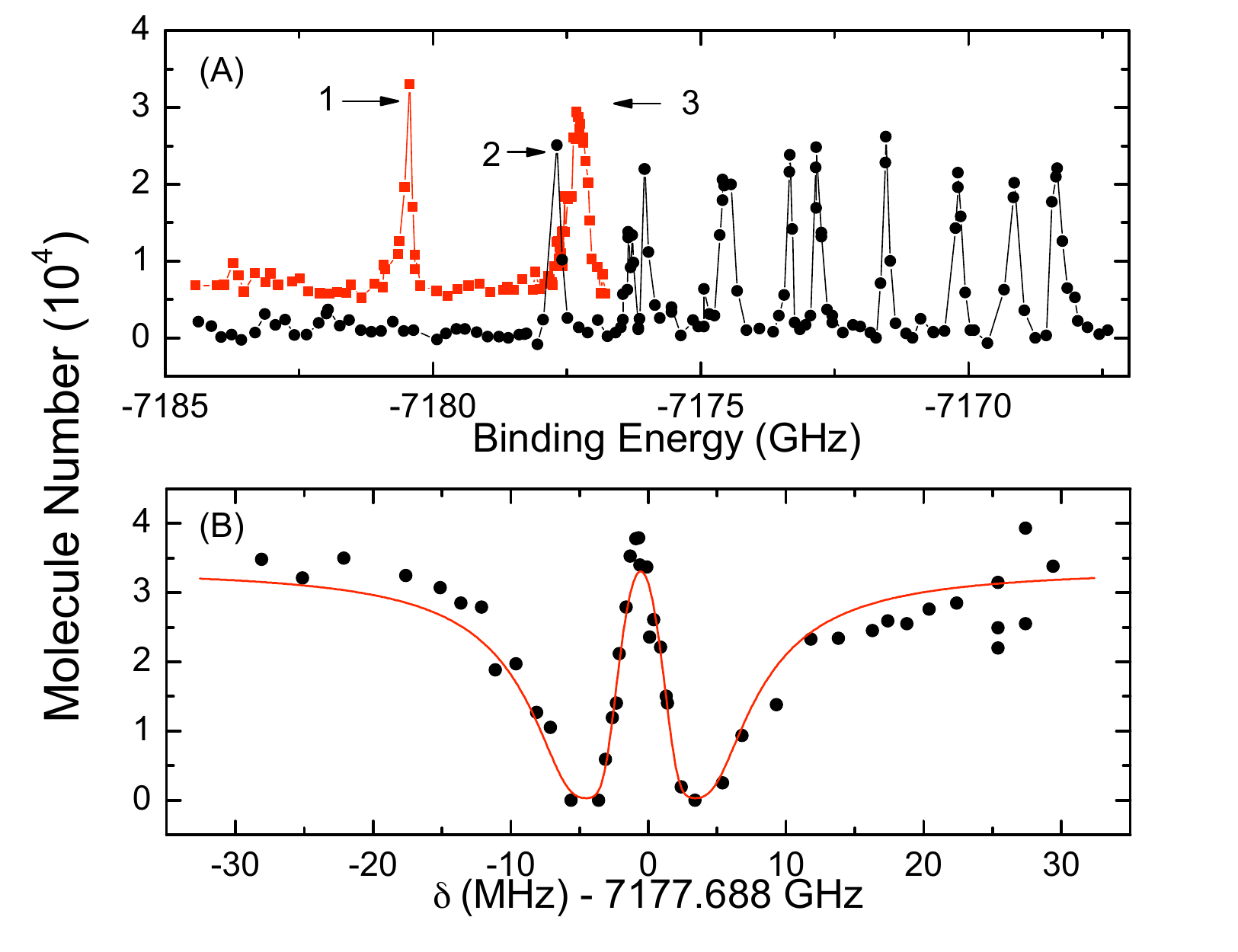}
   \caption{ \doublespacing The $v=0$ ground-state level of the triplet electronic ground potential, $a^3\Sigma$.
   A. Hyperfine and rotational states of the $a^3\Sigma$ $v=0$ ground-state molecule at a magnetic field of 546.94 G, observed using two-photon spectroscopy 
   and scanning the down leg ($\Omega_2$) frequency. The
measured number of Feshbach molecules is plotted as a function of
the frequency difference of the two laser fields. We show two sets
of data, vertically offset for clarity, obtained using two different intermediate states, which are
hyperfine and rotational states of the $v'=10$ level of the
electronically excited $2^3\Sigma$ potential. Peaks labelled 1 and 2 correspond to hyperfine 
states in the 
rotational ground-state, while peak 3 corresponds to a rotationally excited state.
B. We precisely determine
the energy and the transition dipole moments for individual states
using the two-photon spectroscopy where we scan the up leg ($\Omega_1$) frequency.
The measured number of Feshbach molecules is plotted as a function
of the two-photon detuning. The dark resonance shown here
is for the triplet rovibrational ground state corresponding to peak 2 in (A). 
\label{fig:2}}
\end{center}
\vspace{-0.05\textwidth}
\end{figure}

\newpage
\begin{figure}[t]
\begin{center}
   \includegraphics[totalheight=11cm,width=10cm]{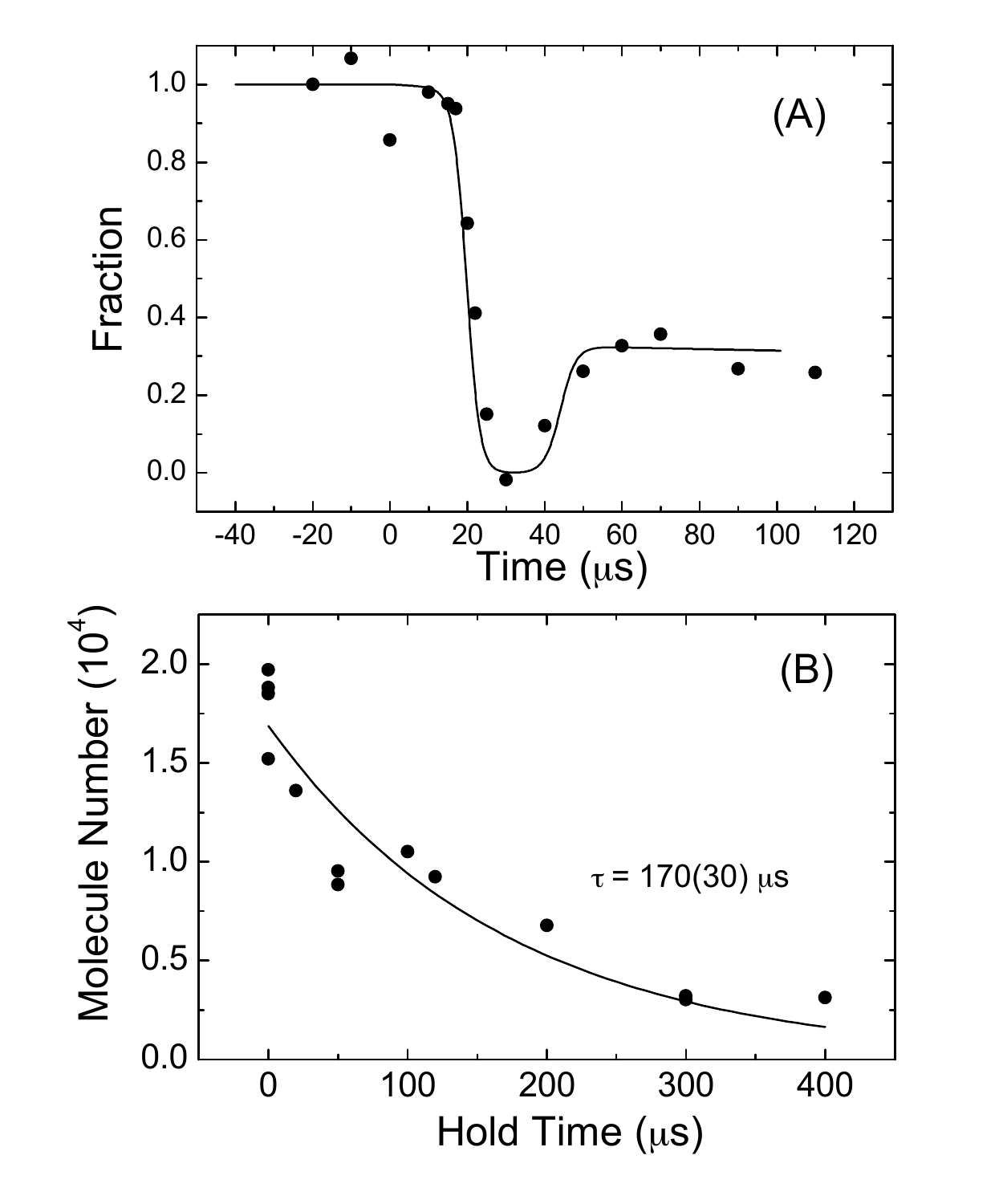}
\caption{\doublespacing Time evolution of the initial state population during STIRAP state transfer and a measurement
of the triplet rovibrational ground-state molecule lifetime.  A. Here we monitor the Feshbach molecule population as we apply the STIRAP
   pulse sequence. Weakly-bound Feshbach molecules are coherently
   transferred into the triplet rovibrational ground state after a 25 $\mu$s one-way STIRAP pulse
   sequence.
   The measured population completely disappears since the 
   deeply-bound molecules are dark to the imaging light. After a 10 $\mu$s
   hold, we then perform the reversed STIRAP pulse sequence that coherently transfers the
   ground-state molecules back to Feshbach molecules. The molecule number after the 
   roundtrip STIRAP is
   $1.8\cdot 10^4$. Assuming equal transfer efficiency for the two STIRAP sequences, we obtain one-way transfer efficiency of 56\% and an absolute number of triplet rovibrational ground-state polar molecules of $3.2\cdot 10^4$. B. We measure the triplet $v=0$ lifetime by varying the hold 
   time after one-way
   STIRAP before transferring them back to Feshbach molecules for imaging.
   The lifetime is measured to be 170(30) $\mu$s.
     \label{fig:3}}
\end{center}
\vspace{-0.05\textwidth}
\end{figure}

\newpage
\begin{figure}[t]
\begin{center}
   \includegraphics[totalheight=10cm,width=13cm]{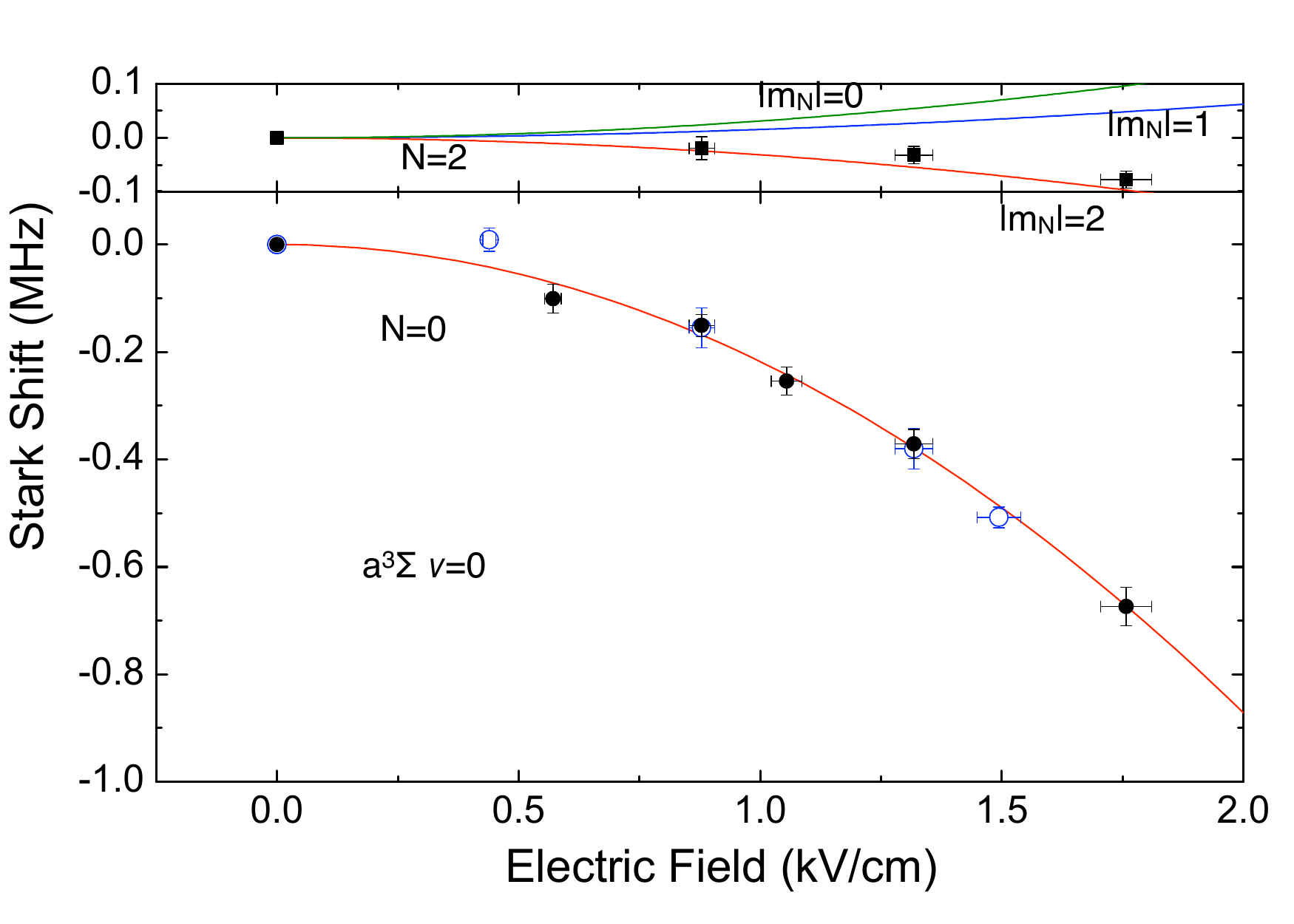}
   \caption{ \doublespacing Stark spectroscopy of the triplet $v=0$ molecules. 
   Stark shifts of the lowest three
   states in the triplet $v=0$ manifold in Fig. 2A are measured for a DC electric field in
   the range from 0 to 2 kV/cm. The bottom panel shows the Stark shifts of the two lowest
   energy states which are $N=0$. Fits to the shifts of peak 1 (solid circles) and
   peak 2 (open circles) yield an electric dipole moment and statistical error bar
   of 0.052106(2) D and 0.052299(8) D, respectively.
   The main systematic error in the dipole moment measurement comes from a 3\% uncertainty of
   the electric field. A combined fit for shifts of
   peak 1 and peak 2 is shown as the solid curve in the bottom panel. 
   With the electric field uncertainty, we obtain an electric dipole moment of 0.052(2) D. The top panel shows the Stark shift of peak 3 (squares) and the expected $N=2$ curves 
   calculated for an electric dipole moment of 0.052 D and different $|m_N|$ projections. 
        \label{fig:4}}
\end{center}
\vspace{-0.05\textwidth}
\end{figure}

\newpage
\begin{figure}[t]
\begin{center}
   \includegraphics[totalheight=12cm,width=16cm]{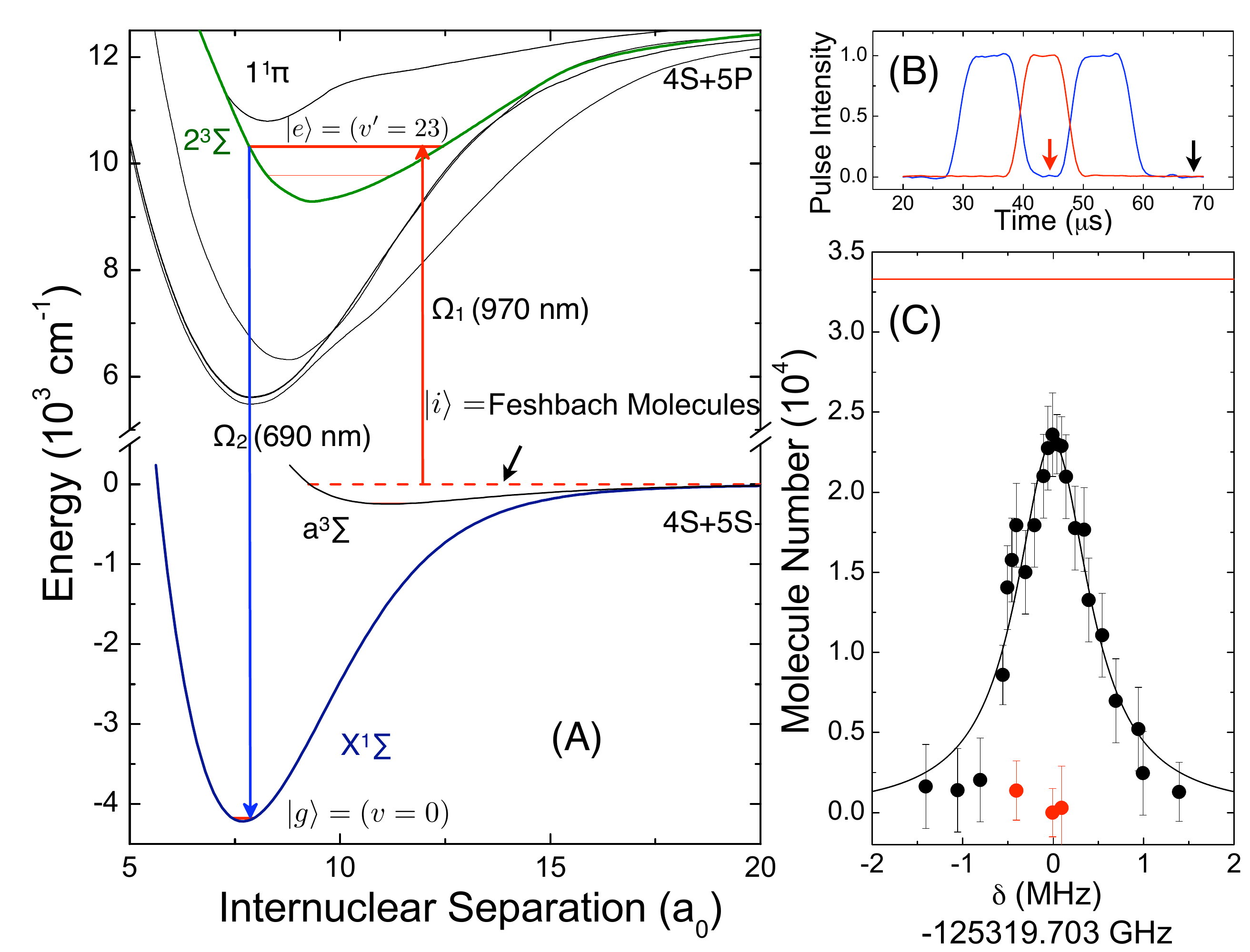}
   \caption{ \doublespacing Two-photon coherent state transfer from 
   weakly-bound Feshbach molecules $|i\rangle$ to the absolute molecular ground state $|g\rangle$
   ($v=0$, $N=0$ of $X^1\Sigma$). 
A. Transfer scheme. 
Here, the intermediate state $|e\rangle$ is the $v'=23$ level of the 
 $\Omega=1$ component of the electronically
excited $2^3\Sigma$ potential.  The chosen intermediate state
lies just below the $1^1\Pi$ excited electronic potential, which provides the necessary 
triplet-singlet spin
mixing to transfer predominantly triplet character Feshbach molecules to the rovibrational ground 
state of the singlet electronic ground potential, $X^1\Sigma$.
The vertical arrows are placed at the respective Condon 
points of the up and down transitions. The intermediate state has favorable
Franck-Condon factors for both the up leg ($|i\rangle$ to $|e\rangle$) and
the down leg ($|e\rangle$ to $|g\rangle$).
        \label{fig:5}}
\end{center}
\vspace{-0.05\textwidth}
\end{figure}

\clearpage
\noindent 
B. Normalized Raman laser intensities vs time for the roundtrip STIRAP pulse sequence.
We performed a 4$\mu$s STIRAP 
transfer each way using a maximum Rabi frequency of 2$\pi \cdot 7$ MHz 
for the downward transition (blue line) and a maximum Rabi frequency of 2$\pi \cdot 4$ MHz
for the upward transition (red line).
C. STIRAP lineshape. The number of Feshbach molecules returned after a roundtrip 
STIRAP transfer is plotted as a function of the two-photon Raman laser detuning. The roundtrip
data were taken at the time indicated by the black arrow in (B). 
The red data points show the Feshbach molecule number 
when only one-way STIRAP is performed (at the time indicated by the red arrow in (B)), where
all Feshbach molecules are transferred to the ground state and are dark to the imaging light.
The initial Feshbach molecule number 
is $3.3(4)\cdot 10^4$ (red solid line) and the number after roundtrip STIRAP 
is $2.3\cdot 10^4$. The roundtrip
efficiency is 69\%, which suggests the one-way transfer efficiency is 83\% and 
the number of the absolute 
ground-state polar molecules is $2.7\cdot 10^4$ .

\newpage
\begin{figure}[t]
\begin{center}
   \includegraphics[totalheight=10cm,width=13cm]{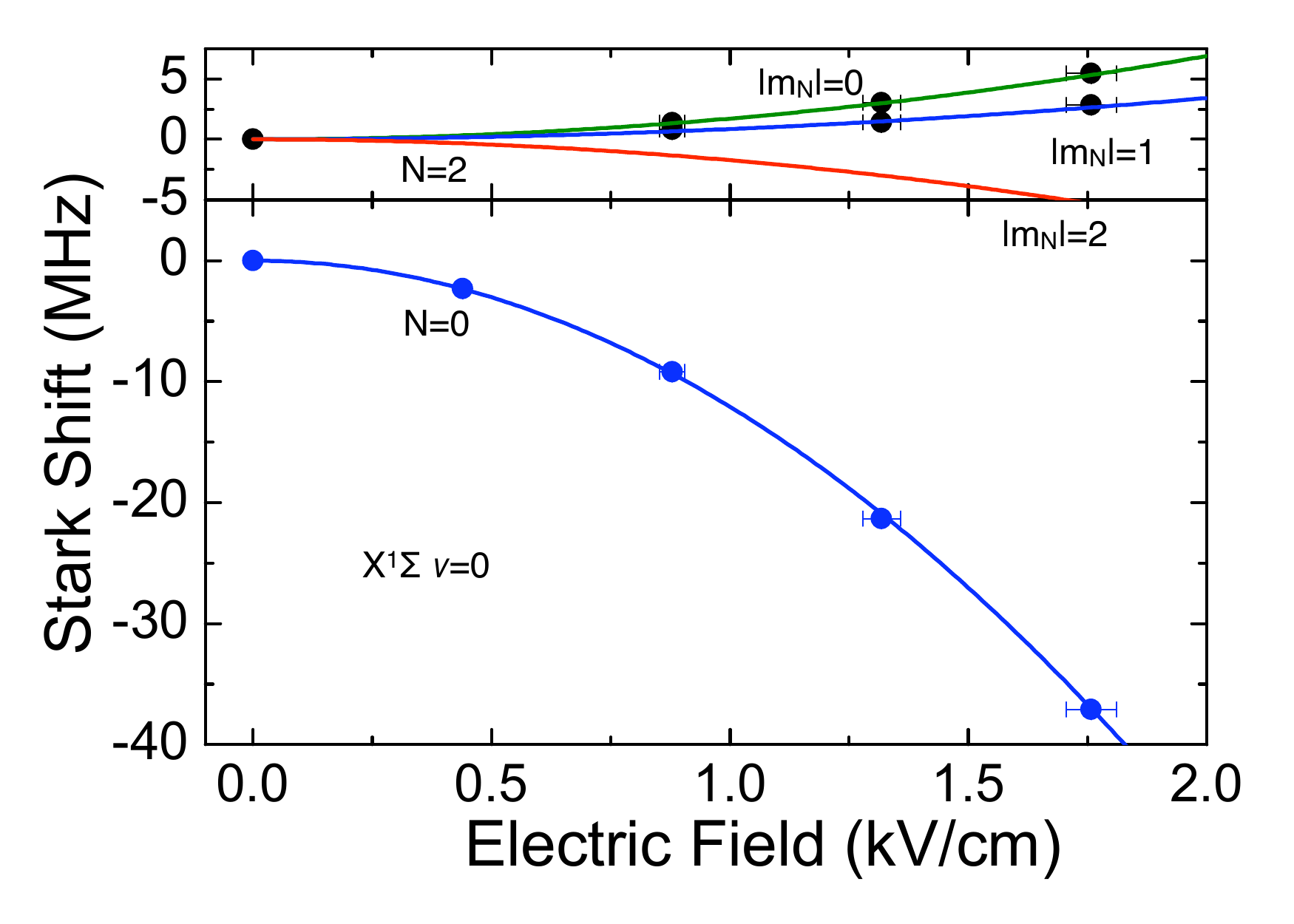}
   \caption{ \doublespacing Stark spectroscopy of the singlet $v=0$ molecules. 
   The bottom panel shows the Stark shift of the rovibrational ground-state of the singlet 
   potential ($v=0, N=0$ of $X^1\Sigma$), and the top panel
   shows the shift of the $v=0, N=2$ state. 
    The systematic error in the applied electric field is 3\%.
   The level difference between $N=0$ and $N=2$ is 6.6836 GHz, which yields
   a rotational constant $B$ of 1.1139(1) GHz. Given the measured $B$, the fit of the
   Stark shift (line in lower panel) gives a permanent electric dipole moment of 0.566(17) D. 
   The theory curves for $N=2$ for different $|m_N|$ projections (lines in upper panel)
   are calculated using the measured $B$ and the dipole moment derived from the 
   $N=0$ fit.
              \label{fig:6}}
\end{center}
\vspace{-0.05\textwidth}
\end{figure}

\clearpage

\end{document}